\documentclass[twocolumn]{aastex62}
\usepackage{CJK}
\usepackage{epsfig}
\usepackage{natbib}
\usepackage{amsmath}
\usepackage{verbatim}
\definecolor{blue}{RGB}{50, 80, 255}
\bibpunct{(}{)}{;}{a}{}{,}      % to give references the ApJ format
\shorttitle{He {\small I} $\lambda$5876 in White Dwarfs}
\shortauthors{B. Klein et al.}

\begin{document}
\begin{CJK*}{UTF8}{gbsn}

\title{Atmospheric Temperature Inversions and He {\small I} $\lambda$5876 Core Profile Structure in White Dwarfs}

\correspondingauthor{Beth Klein}
\email{kleinb@astro.ucla.edu}

\author[0000-0001-5854-675X]{Beth Klein}
\affiliation{Department of Physics and Astronomy, University of California, Los Angeles, CA 90095-1562, USA}

\author[0000-0002-9632-1436]{Simon Blouin}
\affiliation{Los Alamos National Laboratory, P.O. Box 1663, Mail Stop P365, Los Alamos, NM 87545, USA}

\author[0000-0002-5320-4424]{Diego Romani}
\affiliation{Department of Physics, University of California, Santa Barbara, CA 93106-9530, USA}

\author[0000-0001-6809-3045]{B. Zuckerman}
\affiliation{Department of Physics and Astronomy, University of California, Los Angeles, CA 90095-1562, USA}

\author[0000-0001-9834-7579]{Carl Melis}
\affiliation{Center for Astrophysics and Space Sciences, University of California, San Diego, CA 92093-0424, USA}

\author[0000-0002-8808-4282]{Siyi Xu (许\CJKfamily{bsmi}偲\CJKfamily{gbsn}艺)}
\affiliation{NSF's NOIRLab/Gemini Observatory, 670 N. A'ohoku Place, Hilo, Hawaii, 96720, USA}

\author[0000-0003-4609-4500]{P. Dufour}
\affiliation{Institut de Recherche sur les Exoplan\`etes (iREx), Universit\'e de Montr\'eal, Montr\'eal, QC H3C 3J7, Canada}
\affiliation{D\'epartement de physique, Universit\'e de Montr\'eal, Montr\'eal, QC H3C 3J7, Canada}

\author{C. Genest-Beaulieu}
\affiliation{D\'epartement de physique, Universit\'e de Montr\'eal, Montr\'eal, QC H3C 3J7, Canada}

\author[0000-0002-2384-1326]{A. B\'edard}
\affiliation{D\'epartement de physique, Universit\'e de Montr\'eal, Montr\'eal, QC H3C 3J7, Canada}

\author{M. Jura}
\altaffiliation{Deceased}
\affiliation{Department of Physics and Astronomy, University of California, Los Angeles, CA 90095-1562, USA}

\begin{abstract}
We report distinctive core profiles in the strongest optical helium line, He {\small I} $\lambda$5876{\AA}, from high-resolution high-sensitivity observations of spectral type DB white dwarfs. By analyzing a sample of 40 stars from Keck/HIRES and VLT/UVES, we find the core appearance to be related to the degree of hydrogen and heavy element content in the atmosphere. New Ca K-line measurements or upper limits are reported for about half the sample stars. He {\small I} $\lambda$5876{\AA} emission cores with a self-reversed central component are present for those stars with relatively low hydrogen abundance, as well as relatively low atmospheric heavy element pollution. This self-reversed structure disappears for stars with higher degrees of pollution and/or hydrogen abundance, giving way to a single absorption core. From our model atmospheres, we show that the self-reversed emission cores can be explained by temperature inversions in the upper atmosphere. We propose that the transition to a single absorption core is due to the additional opacity from hydrogen and heavy elements that inhibits the temperature inversions. Our current models do not exactly match the effective temperature range of the phenomenon nor the amplitude of the self-reversed structure, which is possibly a result of missing physics such as 3D treatment, convective overshoot, and/or non-LTE effects. The He {\small I} $\lambda$5876{\AA} line structure may prove to be a useful new diagnostic for calibrating temperature profiles in DB atmosphere models.  \\

\end{abstract}

%\keywords{Stars: abundances --  white dwarfs -- atmospheres}

\section{Introduction \label{sec:intro}}
The presence of elements heavier than helium in the atmospheres of many single white dwarfs (WDs) has garnered increasing interest over the past decade mainly due to the recognition that such stars have accreted material from their extant planetary systems \citep[e.g.~reviews by][]{jurayoung2014, veras2016, farihi2016, zuckermanyoung2018}.   
We now know that white dwarfs ``polluted'' by heavy (high-Z\footnote{We use the terms `heavy elements' and `high-Z' to refer to elements heavier than hydrogen or helium.}) elements are powerful observational tools that provide insights into aspects of exoplanetary systems that are elusive or inaccessible via other techniques.  For example, pollution in a single WD suggests the existence of an asteroidal or cometary debris belt and at least one major planet with wide semi-major axis \citep{veras2013, frewenhansen2014, mustill2014, mustill2018}, both of which have significant observational limitations in main sequence systems.   Moreover, high-resolution high-sensitivity spectroscopy of these stars has revealed a rich and informative laboratory previously hidden in low resolution observations, as the elemental constituents of accreted bodies can be measured at an extraordinary level of detail and precision. In particular, it has opened an invaluable window to detailed exoplanet composition measurements beginning with works that include \citet{zuckerman2007}, \citet{klein2010, klein2011}, \citet{dufour2010}, \citet{vennes2010, vennes2011}, \citet{melis2011},  \citet{farihi2011gd61}, \citet{jura2012cos}, and \citet{gaensicke2012cos}.

Meanwhile, the presence and origin of hydrogen in the atmospheres of helium-dominated white dwarfs is a related subject of ongoing investigation. Throughout the evolution of a WD, different mechanisms can raise or lower the hydrogen abundance at its photosphere, a phenomenon known as spectral evolution. The exact roles of transport mechanisms within the WD envelope \citep{macdonald1991,rolland2018,rolland2020} and of accretion from the interstellar medium or from water-bearing planetesimals \citep{macdonald1991,bergeron2011,veras2014,koesterkepler2015,raddi2015,gentilefusillo2017} are still being debated.

The conditions of a WD atmosphere affect the visibility of pollution in resulting spectra.  The dominant background element $-$ either hydrogen or helium $-$ sets the stage with opacity. Compared to hydrogen, the lower opacity of a helium-dominated atmosphere produces relatively larger, thus more easily detectable, absorption features for a given high-Z abundance.  Additionally, the effective temperature ($T_{\rm eff}$) has a significant effect, as many important optical lines of major and minor elements disappear at either (or both) of the high and low extremes of $T_{\rm eff}$ as the element ionization state changes.  Due to the combination of these influences, intermediate temperature helium-dominated WDs are often the richest systems for observing pollution and conducting exoplanet composition analyses.

The DB class is defined as those WDs for which the strongest lines in optical spectra are from He {\small I} \citep{luyten1952, sion1983}. These are hydrogen-deficient WDs in the range of $T_{\rm eff}$ $\sim$10,000$-$40,000\,K.  Below $\sim$12,000\,K neutral helium lines become very weak and essentially disappear at cooler temperatures, while above $\sim$40,000\,K lines of He {\small II} begin to dominate, changing the spectral classification type to DO.  If hydrogen lines are also visible, but weaker than the helium lines, then an ``A'' is added to the spectral type (e.g.~DBA), and if high-Z elements are detected a ``Z'' is added (DBZ or DBAZ). In DBs the strongest optical He {\small I} line is at $\lambda$5875.615 {\AA} in air (hereafter, He5876).

Helium absorption line strengths and profiles in white dwarf optical spectra have been widely used to measure the fundamental stellar parameters of effective temperature, $T_{\rm eff}$,  and gravity (log $g$), by fitting the helium lines with synthetic spectra \citep{koester1981, beauchamp1996, eisenstein2006, voss2007, bergeron2011, koesterkepler2015, rolland2018}. Alternatively, $T_{\rm eff}$ and log $g$ can be derived from fitting the spectral energy distribution utilizing broadband photometry and parallaxes \citep[e.g.][]{bergeron1997, bedard2017}. Thanks to the {\it Gaia} satellite mission \citep{gaiacollaboration2016}, with its photometry and parallax measurements over the whole sky, we now have large catalogs of WDs that include such photometric fits \citep{dufourMWDD, gentilefusillo2019}.  Comparisons between these photometric versus spectroscopic fits of $T_\textrm{eff}$ and log $g$ have been comprehensively investigated, and while there is good overall agreement in derived parameters from the two methods, some systematic discrepancies remain \citep{tremblay2019, genestbergeron2019, bergeron2019}.  

Another outstanding problem in modeling polluted WDs are some disagreements in high-Z abundances derived from UV and optical data of the same element \citep{jura2012cos, gaensicke2012cos, koester2014da, melisdufour2017, xudusty2019}.  While more work is needed to solve these UV-optical discrepancies, it may likely have to do with uncertain atomic data (such as oscillator strengths and Stark widths/shifts) and/or chemical stratification \citep{vennes2011, gaensicke2012cos}.
Meanwhile, an advance in resolving the differences between hydrogen abundances in DBs derived from H$\alpha$ and Ly$\alpha$ was made by incorporating broadening by neutral helium into the models \citep{gaensickelyalpha2018, allard2020}.   Some other potentially important effects under consideration are departures from local thermodynamic equilibrium \citep[non-LTE effects, e.g.][]{napiwotzki1997,hubeny1999}, 3-D models \citep{tremblay2013,cukanovaite2018}, convective overshoot \citep{tremblayludwig2017,kupka2018,cunningham2019}, and thermohaline mixing \citep{deal2013,koester2015,bauerbildsten2018}.

As model atmosphere theories and calculations continue to improve and incorporate additional physics related to atmospheric structure, it is desirable to acquire new observational diagnostic tools to guide the way. This paper is a step in that direction. Here we present the serendipitous discovery of distinctive core profiles of the most prominent optical helium line, He5876 in DB WDs, and we demonstrate a connection between atmospheric hydrogen and/or pollution and the appearance of the line core shape. 

The paper is organized as follows:~in Section \ref{sec:sample} we describe the observations, data processing, and how we arrived at our sample of DB stars. Section \ref{sec:pollution} discusses details of our measurements of heavy element pollution and derived abundances.  A categorization and analysis of the sample stars according to their hydrogen abundance and pollution from high-Z elements is given in Section \ref{sec:profiles}.  In Section \ref{sec:model} we offer our modeling insights about the presence (or lack) of the He5876 core-inversion feature. Conclusions are given in Section \ref{sec:conclusions}.

\startlongtable
\begin{deluxetable*}{llclcrrrcrrcl}
\tablecaption{DB White Dwarf Parameters \label{tab:params} }
\tabletypesize{\footnotesize}
%\begin{center}
%\begin{tabular}{llcrcrrrrcl}
%\hline 
%\hline
%\tablenotemark{a}
\tablehead{
    \colhead{WD \#} & \colhead{Name} & \colhead{Instru-} & \colhead{$G$} & \colhead{He5876} &  \colhead{$T_\textrm{eff}$}  & \colhead{log $g$}  & \colhead{[H/He]} & \colhead{atm.} & \colhead{$EW$ CaK}  & \colhead{[Ca/He] } &  \colhead{$EW$,abund}   \\
\colhead{}	  & 	\colhead{}	 & \colhead{ment}  & \colhead{(mag)} & \colhead{core} & \colhead{(K)}  &  \colhead{} &  \colhead{}   & \colhead{ref.} & \colhead{(m{\AA})}  & \colhead{}  & \colhead{ref.} 
}
\startdata
0002+729	&	GD 408	&	HIRES	&	14.29	&	inv	&	14410	&	8.27	&	$-$5.95	&	R	&	175 $\pm$ 10	&	$-$9.59	&	(1*)	\\
0017+136	&	Feige 4	&	HIRES	&	15.36	&	abs	&	18130	&	8.08	&	$-$4.63	&	R	&	$<$ 10	&	$<$ $-$9.1	&	(1)	\\
0100$-$068	&	G270-124	&	HIRES	&	13.91	&	abs	&	19820	&	8.06	&	$-$5.14	&	R	&	27 $\pm$ 4	&	$-$8.08	&	(1*)	\\
0110$-$565	&	HE 0110-5630	&	UVES	&	15.79	&	abs	&	18483	&	8.12	&	$-$4.18	&	V	&	130	&	$-$7.9	&	(3)	\\
0300$-$013	&	GD 40	&	both	&	15.51	&	abs	&	14620	&	7.99	&	$-$6.14	&	R	&	2500 $\pm$ 200	&	$-$6.88	&	(4)	\\
0308$-$565	&	BPM 17088	&	UVES	&	14.12	&	abs	&	22840	&	8.07	&	$<$ $-$4.82	&	R	&	$<$ 15	&	$<$ $-$7.8	&	(1)	\\
0435+410	&	GD 61	&	HIRES	&	14.84	&	abs	&	16790	&	8.18	&	$-$4.21	&	R	&	173 $\pm$ 12	&	$-$7.9	&	(1), (5)	\\
0437+138	&	LP475-242	&	HIRES	&	14.93	&	inv	&	15120	&	8.25	&	$-$4.68	&	R	&	40 $\pm$ 8	&	$-$9.2	&	(6)	\\
0503+147	&	KUV 05034+1445	&	HIRES	&	14.12	&	inv	&	15640	&	8.09	&	$-$5.46	&	R	&	$<$ 8	&	$<$ $-$10.7	&	(1)	\\
0615$-$591	&	L182-61	&	UVES	&	13.97	&	inv	&	15770	&	8.04	&	$<$ $-$6.32	&	R	&	$<$ 20	&	$<$ $-$10.3	&	(1)	\\
0716+404	&	GD 85	&	HIRES	&	14.89	&	inv	&	17150	&	8.08	&	$<$ $-$5.99	&	R	&	53 $\pm$ 5	&	$-$8.80	&	(1*)	\\
0842+231	&	Ton 345	&	HIRES	&	15.90	&	abs	&	19780	&	8.18	&	$-$5.10	&	W	&	350 $\pm$ 20	&	$-$5.95	&	(1), (7)	\\
1011+570	&	GD 303	&	HIRES	&	14.62	&	abs	&	17610	&	8.16	&	$<$ $-$5.34	&	R	&	126 $\pm$ 8	&	$-$7.8	&	(1*)	\\
1046$-$017	&	GD 124	&	both	&	15.77	&	inv	&	14620	&	8.15	&	$<$ $-$6.46	&	R	&	$<$ 9	&	$<$ $-$10.9	&	(2)	\\
1056+345	&	G119-47	&	HIRES	&	15.53	&	abs	&	12440	&	8.23	&	$-$5.33	&	R	&	$<$ 8	&	$<$ $-$11.9	&	(2)	\\
1144$-$084	&	PG 1144-085	&	UVES	&	16.09	&	inv	&	15730	&	8.06	&	$<$ $-$6.32	&	R	&	$<$ 33	&	$<$ $-$10.1	&	(1)	\\
1252$-$289	&	EC 12522-2855	&	UVES	&	15.89	&	abs	&	21880	&	8.03	&	$<$ $-$4.82	&	R	&	$<$ 23	&	$<$ $-$7.8	&	(1)	\\
1326$-$037	&	PG 1326-037	&	UVES	&	15.76	&	abs	&	19950	&	8.03	&	$<$ $-$4.81	&	R	&	$<$ 27	&	$<$ $-$8.0	&	(1)	\\
1333+487	&	GD 325	&	HIRES	&	14.09	&	inv	&	15420	&	8.01	&	$-$6.37	&	R	&	$<$ 10 	&	$<$ $-$10.6	&	(1)	\\
1403$-$010	&	G64-43	&	both	&	15.71	&	inv	&	15420	&	8.10	&	$-$6.08	&	R	&	$<$ 10	&	$<$ $-$10.6	&	(2)	\\
1411+218	&	PG 1411+219	&	HIRES	&	14.43	&	inv	&	14970	&	8.02	&	$-$6.26	&	R	&	$<$ 5	&	$<$ $-$10.8	&	(2)	\\
1425+540	&	G200-39	&	HIRES	&	15.04	&	abs	&	14410	&	7.89	&	$-$4.26	&	R	&	235 $\pm$ 15	&	$-$9.3	&	(1), (8)	\\
1459+821	&	G256-18	&	HIRES	&	14.83	&	inv	&	16020	&	8.08	&	$<$ $-$6.28	&	R	&	$<$ 7	&	$<$ $-$10.7	&	(2)	\\
1542+182	&	GD 190	&	UVES	&	14.67	&	abs	&	22620	&	8.04	&	$<$ $-$4.84	&	R	&	$<$ 22	&	$<$ $-$7.6	&	(1)	\\
1644+198	&	PG 1644+199	&	HIRES	&	15.07	&	inv	&	15210	&	8.14	&	$-$5.68	&	R	&	17 $\pm$ 1	&	$-$10.3	&	(2)	\\
1645+325	&	GD 358	&	HIRES	&	13.59	&	abs	&	24940	&	7.92	&	$<$ $-$4.58	&	R	&	$<$ 8	&	$<$ $-$7.7	&	(1)	\\
1709+230	&	GD 205	&	both	&	14.84	&	abs	&	19590	&	8.08	&	$-$4.07	&	R	&	45 $\pm$ 4	&	$-$7.9	&	(2)	\\
1822+410	&	GD 378	&	HIRES	&	14.28	&	abs	&	16230	&	8.00	&	$-$4.45	&	R	&	160 $\pm$ 10	&	$-$8.3	&	(1*)	\\
1916$-$362	&	WD 1916-362	&	HIRES	&	13.58	&	abs	&	23610	&	8.10	&	$-$4.22	&	R	&	$<$ 10	&	$<$ $-$7.6	&	(1)	\\
1917$-$074	&	LAWD 74	&	both	&	12.27	&	abs	&	10195	&	8.0	&	$-$5.16	&	V	&	$<$ 6	&	$<$ $-$12.5	&	(1)	\\
1940+374	&	EGGR 133	&	HIRES	&	14.49	&	inv	&	16850	&	8.07	&	$-$5.97	&	R	&	$<$ 5	&	$<$ $-$10.1	&	(2)	\\
2129+000	&	G26-10	&	HIRES	&	14.66	&	inv	&	14350	&	8.25	&	$<$ $-$6.49	&	R	&	$<$ 7	&	$<$ $-$11.1	&	(2)	\\
2144$-$079	&	G26-31	&	both	&	14.79	&	inv	&	16340	&	8.18	&	$<$ $-$6.22	&	R	&	132 $\pm$ 7	&	$-$8.6	&	(2)	\\
2154$-$437	&	LAWD 90	&	UVES	&	14.64	&	abs	&	16734	&	8.02	&	$-$4.78	&	V	&	$<$ 20	&	$<$ $-$9.6	&	(1)	\\
2222+683	&	G241-6	&	HIRES	&	15.67	&	abs	&	14920	&	8.00	&	$-$6.43	&	R	&	960 $\pm$ 96	&	$-$7.25	&	(2)	\\
2253$-$062	&	GD 243	&	UVES	&	15.05	&	abs	&	17190	&	8.07	&	$-$4.35	&	R	&	$<$ 21	&	$<$ $-$9.1	&	(1)	\\
2316$-$173	&	L791-40	&	UVES	&	14.08	&	abs	&	10868	&	8.0	&	$-$5.27	&	V	&	$<$ 20	&	$<$ $-$12.0	&	(1)	\\
2328+510	&	GD 406	&	HIRES	&	15.08	&	inv	&	14500	&	8.03	&	$-$6.47	&	R	&	$<$ 12	&	$<$ $-$10.8	&	(1)	\\
2334$-$414	&	HE 2334-4127	&	UVES	&	15.16	&	abs	&	18250	&	8.03	&	$-$5.34	&	V	&	$<$ 16	&	$<$ $-$8.9	&	(1)	\\
2354+159	&	PG 2354+159	&	UVES	&	15.75	&	abs	&	24830	&	8.15	&	$<$ $-$4.59	&	R	&	80	&	$-$8.1	&	(3)	\\
\enddata
%\end{tabular}
%\end{center}
\tablecomments{The appearance of the He5876 core is indicated either as an inversion (inv) shown in Figure \ref{fig:lowZ}, or in absorption (abs) with a representative set shown in Figure \ref{fig:highZ}. Magnitudes are {\it Gaia} $G$. Atmospheric parameters, $T_\textrm{eff}$, log $g$, and the logarithmic abundance of hydrogen relative to helium by number ([H/He]), come from spectroscopic fits by R=\citet{rolland2018}, V=\citet{voss2007}, or W=\citet{wilson2015}, as listed in the column labeled ``atm. ref.". Equivalent widths ($EW$) and upper limits are for the photospheric Ca K-line ($\lambda$ = 3933.663 {\AA}), either newly presented here or taken from the literature as indicated by the ``$EW$,abund ref.'' column (see reference list below).  For the logarithmic abundance of calcium relative to helium by number ([Ca/He]), if the only entry in the $EW$,abund ref. column is (1), then these are new values we calculated by comparing WDs with similar $T_\textrm{eff}$ and [Ca/He] from \citet{zuckerman2010} and scaling the abundance by the ratio of the Ca K-line $EW$ measurements.   If the entry is (1*), we have performed a model abundance fit in this paper.  If there is a second reference next to (1), then just the $EW$ measurement is from this paper and [Ca/He] is from the second reference. (2)$-$(5) have both the $EW$ and [Ca/He] coming from the indicated paper. }
\textbf {References:}  (1) This Paper, (2)  \citet{zuckerman2010}, (3) \citet{koester2005}, (4) \citet{klein2010}, (5) \citet{farihi2013}, (6)  \citet{zuckerman2013}, (7) \citet{wilson2015},  (8) \citet{xu2017}. 
%\tablenotetext{a}{} 
\end{deluxetable*}
%G200-39 and it's CPM companion G200-40 are name-swapped in SIMBAD

%\clearpage
\section{Sample Selection \label{sec:sample}}
During the course of UCLA-based Keck WD programs conducted over the years from 2006$-$2012, spectra covering the He5876 line were obtained with the High Resolution Echelle Spectrograph \citep[HIRES,][]{vogt1994} on the Keck 1 Telescope. Most observations were made with the HIRES blue collimator and a subset of eight of the WDs were also observed with the HIRES red collimator; both of these setups covered the He5876 wavelength region. HIRES data were acquired with the C5 decker, providing a slit width of 1$\farcs$148 and resolving power $R$ = $\lambda \over \Delta\lambda$ $\sim$ 40,000. The wavelength coverage is approximately 3120$-$5950{\AA} in the blue, and 4500$-$9000{\AA} in the red, with small gaps in coverage in-between the 3 CCDs, as well as in-between echelle orders of the red data. Standard IRAF \citep{iraf1986} routines for echelle data reduction followed the procedures described in  \citet{klein2010} and/or used the HIRES software package MAKEE\footnote{\url{https://www.astro.caltech.edu/~tb/makee/}}.

\begin{figure*}
\begin{center}
\includegraphics[width=160mm]{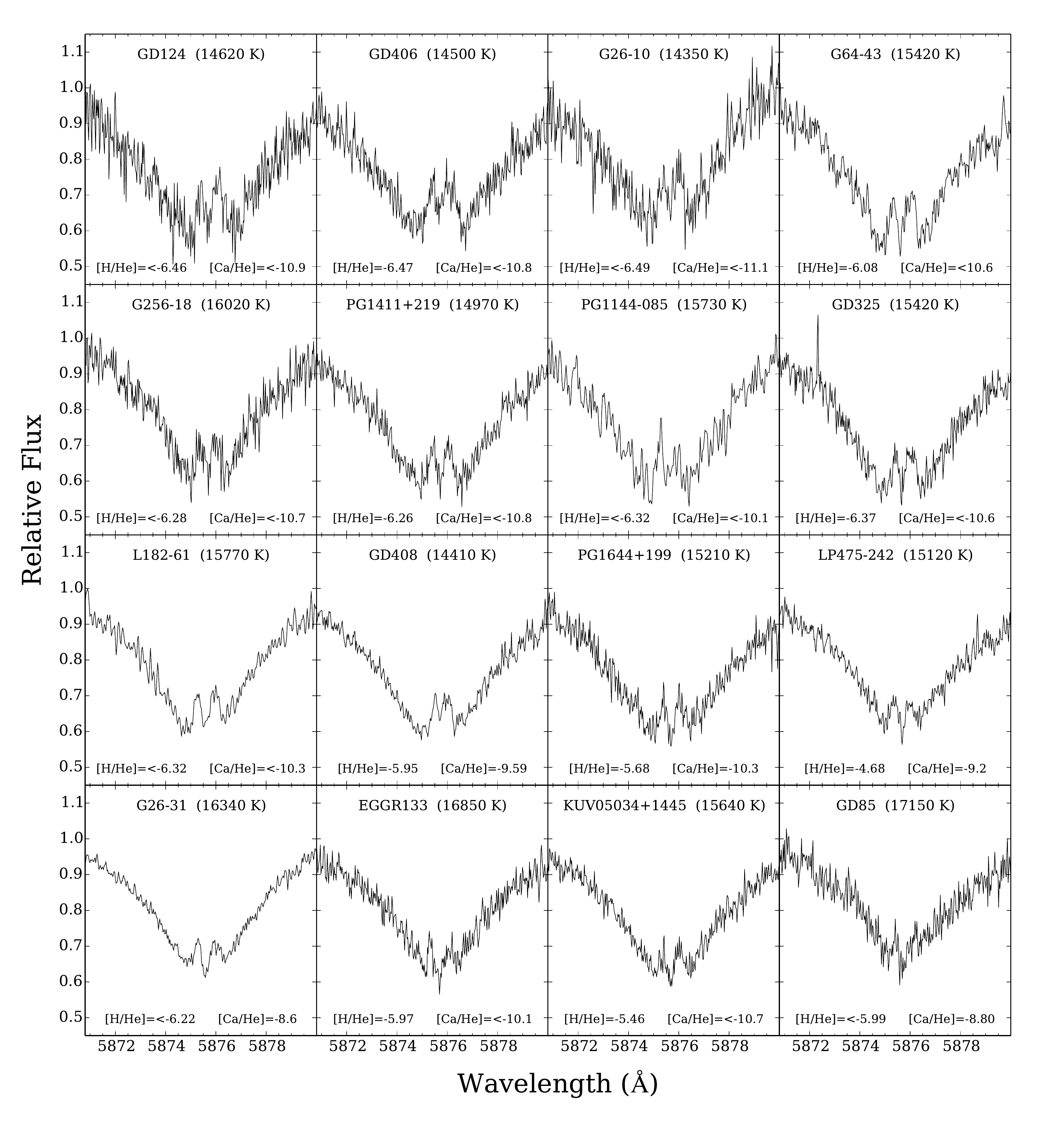}
\caption{Central region of the He5876 line of DB WDs from Table \ref{tab:params} that display a self-reversed inversion core. Spectra are ordered from upper left to lower right by decreasing strength of the inversion feature (see Table \ref{tab:core-data}). Wavelengths are in air and the He5876 rest frame.}
\end{center}
\label{fig:lowZ}
\end{figure*}

\begin{figure*}
\begin{center}
\includegraphics[width=160mm]{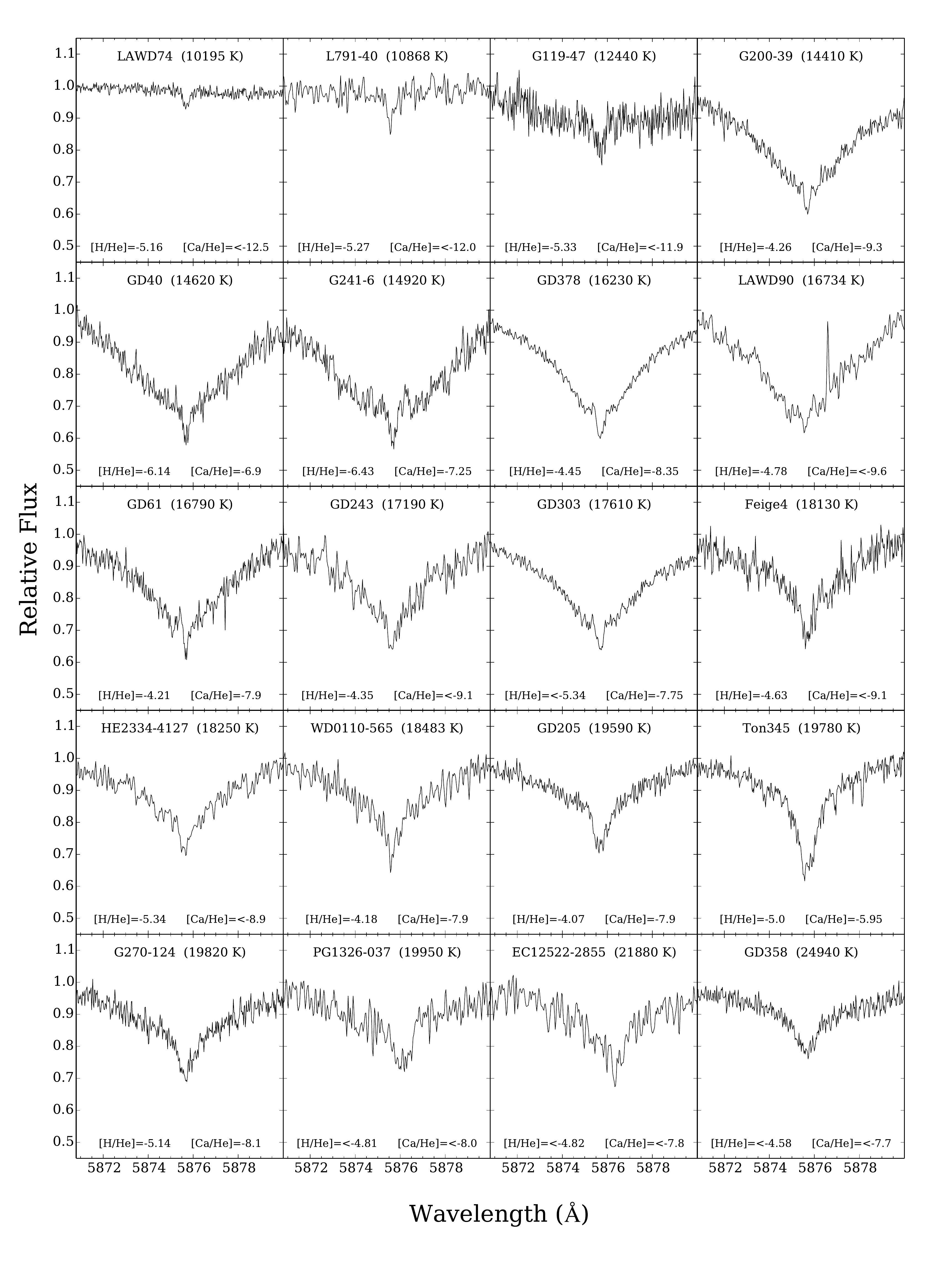}
\caption{Similar to Fig \ref{fig:lowZ}, but for stars with He5876 cores in absorption. Shown here is a representative sub-sample of Table \ref{tab:params} stars over the full temperature range (10,000\,K $<$ $T_\textrm{eff}$ $<$ 25,000\,K), ordered from upper left to lower right by increasing temperature.   Low-level emission components may be present in a few stars (e.g.~G241-6 and GD 378), but their cores are still predominantly in absorption.} 
\label{fig:highZ}
\end{center}
\end{figure*}

None of the target stars were selected with any fore-knowledge of structure in the He5876 line cores. 
Typically our HIRES programs were aimed at searching for, or following up on, atmospheric pollution in WDs, but careful inspection of the helium lines revealed the existence of previously unknown structure in the He5876 line cores for some DBs. A portion of the stars display core emission with a self-reversed central component as shown in Figure \ref{fig:lowZ}, while others have cores in absorption with a variety of shapes and strengths as shown in Figure \ref{fig:highZ}.  Upon further investigation we noticed a trend that the presence of the self-reversed core inversion appears for those stars with little or no atmospheric hydrogen or high-Z elements, while those WDs with evidence for substantial quantities of these elements have He5876 cores predominantly in absorption.   

\begin{figure}
\includegraphics[width=70mm]{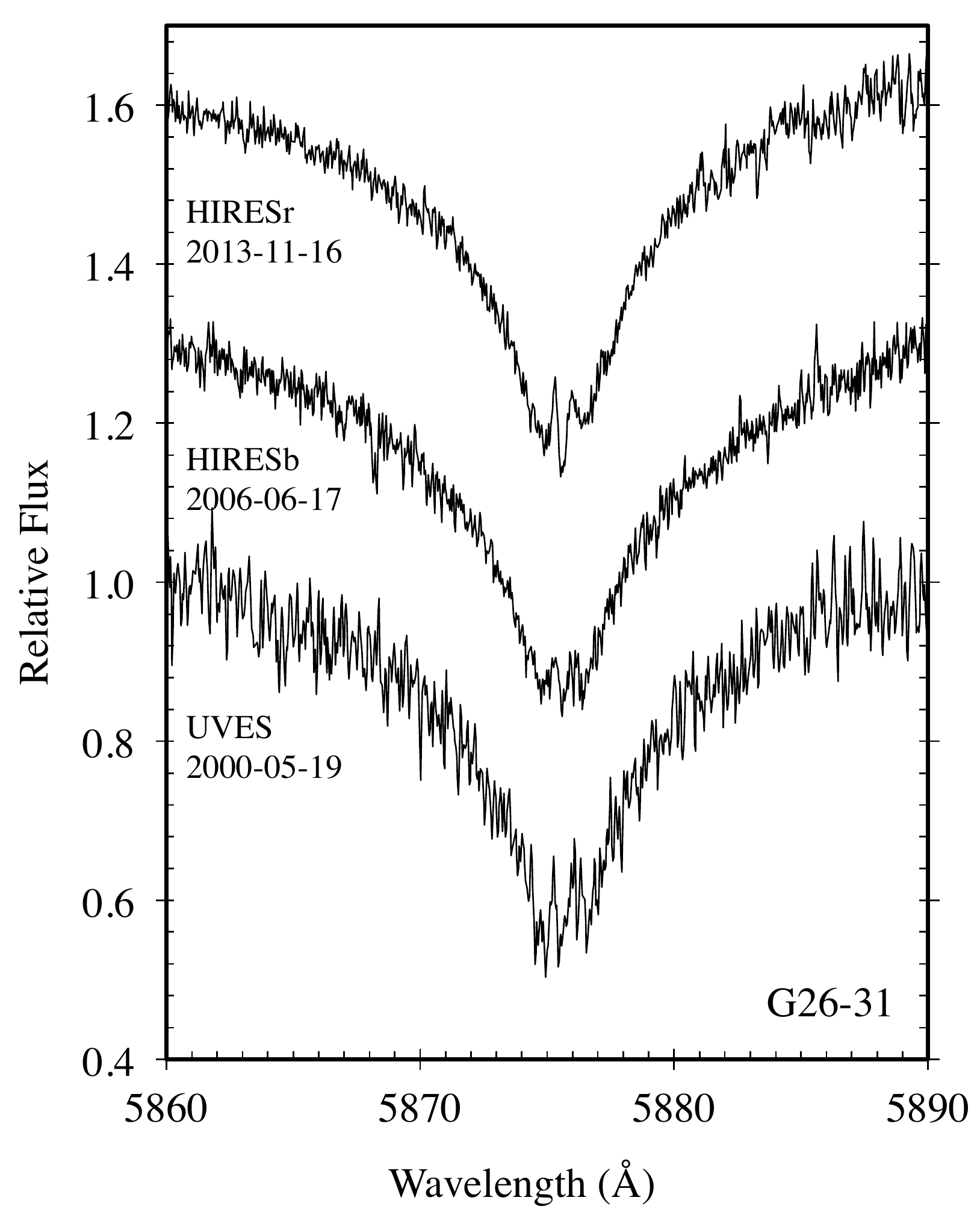}
\caption{Three epochs of G26-31 (= GL837.1 = WD 2144-079) with different instruments.  Spectra are normalized to unity at $\lambda$5860{\AA} and offset by steps of 0.3 in the ordinate.}
\label{fig:g26-31}
\end{figure}

To dismiss any questions of instrumental artifacts and better assess the origin of the line core appearance, we checked the European Southern Observatory (ESO) archival database of high-resolution observations of DBs studied by \citet{voss2007}, i.e.~data originating from the Supernovae Type Ia Progenitor (SPY) Survey \citep{napiwotzki2001}.  Those data were acquired at the Very Large Telescope of ESO with the UV-Visual Echelle Spectrograph (UVES) providing a resolving power $R$$\sim$18,500 and nearly complete wavelength coverage from 3200 {\AA} to 6650 {\AA}.  We found that the He5876 profile in the stars that have observations from both UVES and HIRES instruments (six stars in this study) agree within the noise levels; an example is shown in Figure \ref{fig:g26-31}. Some of the HIRES data and all the UVES data included coverage of H$\alpha$, so we looked for any sign of emission in the H$\alpha$ cores of those stars and found none.

The He5876 core features are relatively weak, and only clearly appear in the most sensitive spectra.  Therefore, we drew our sample from the database of DB WDs observed with either HIRES or UVES by selecting sources with Vmag $<$ 16.0 and a signal-to-noise ratio (SNR) $>$ 25 measured near $\lambda$ = 5869 {\AA}, on the blue wing of the broad He5876 line. This resulted in the set of 40 stars listed in Table \ref{tab:params}.  For ease of display and visual comparison in Figures \ref{fig:lowZ} $-$ \ref{fig:g26-31}, all the spectra were scaled to unity at $\lambda$ = 5869 {\AA} and velocity shifted to the rest frame of He5876, based on the line center positions as measured from the most central core of the line: either the center of an absorption core feature, or the center of a self-reversed emission core.

As mentioned in the Introduction, for a given WD, different studies can result in different values for the fundamental parameters $T_{\rm eff}$, log $g$ and [H/He] ($\equiv$ log n(H)/n(He) ).  Variations come from differences in the spectroscopic versus photometric techniques, ongoing evolution of the models and methods, as well as improvements in observational data.  For example, when dealing with DB WDs, a significant effect on derived atmospheric parameters can occur depending on whether the abundances of hydrogen and heavy elements are included in computing the atmospheric structure \citep[e.g.,][]{dufour2007,dufour2010,blouin2018a,blouin2018b,coutu2019}.

These variations in fundamental parameters will result in variations of the derived [Ca/He] ($\equiv$ log n(Ca)/n(He)) abundances. Nonetheless, in the context of this work they are relatively small perturbations, and do not alter the overall picture of what is happening with the helium lines.  Thus, to create as uniform of a parameter context as possible, we take $T_\textrm{eff}$, log $g$, and [H/He] predominantly from \citet{rolland2018}, since their list of DBs contains nearly all our sample stars, and includes hydrogen abundance measurements or upper limits.  A handful of stars with UVES data do not appear in \citet{rolland2018}, so we use the parameters fit by \citet{voss2007} for those, while the parameters for Ton 345 come from \cite{wilson2015}.  From Table \ref{tab:params} we find that the core inversion phenomenon is only observed over a relatively small range of effective temperatures: 14,000\,K $\lesssim$ $T_{\rm eff}$  $\lesssim$ 17,500\,K.  Nonetheless, we perform our analysis on the entire sample since it is not clear precisely what the upper and lower $T_{\rm eff}$ boundaries are.

\section{Pollution measurements \label{sec:pollution}}
Often the strongest optical line $-$ as defined by the strength of the measured equivalent width (EW) $-$ in a polluted WD is the Ca {\small II} resonance line at $\lambda$ = 3933.663{\AA} (Ca K-line). However, for hydrogen-dominated polluted WDs (DAZ) stars somewhere around $T_{\rm eff}$ $\sim$16,000$-$17,000\,K there is a changeover in the predominant absorption feature from the Ca K-line to Mg {\small II} 4481, which of course has a dependence on the Ca/Mg abundance ratio. We looked for this effect in the DBZ stars, but in our small sample we do not see a cross-over in Ca and Mg line-strengths even up to $\sim$24,000\,K. For the hottest stars displaying optical calcium lines, GD 205, Ton 345, G270$-$124, and PG2354+159, the strength of Mg {\small II} $\lambda$4481 is either comparable to, or weaker than (undetected in G270$-$124 and PG2354+159), that of the Ca K-line.  Thus, we use measurements and upper limits of the Ca K-line for the entire $T_{\rm eff}$ range of our sample.

\begin{figure}
\includegraphics[width=85mm]{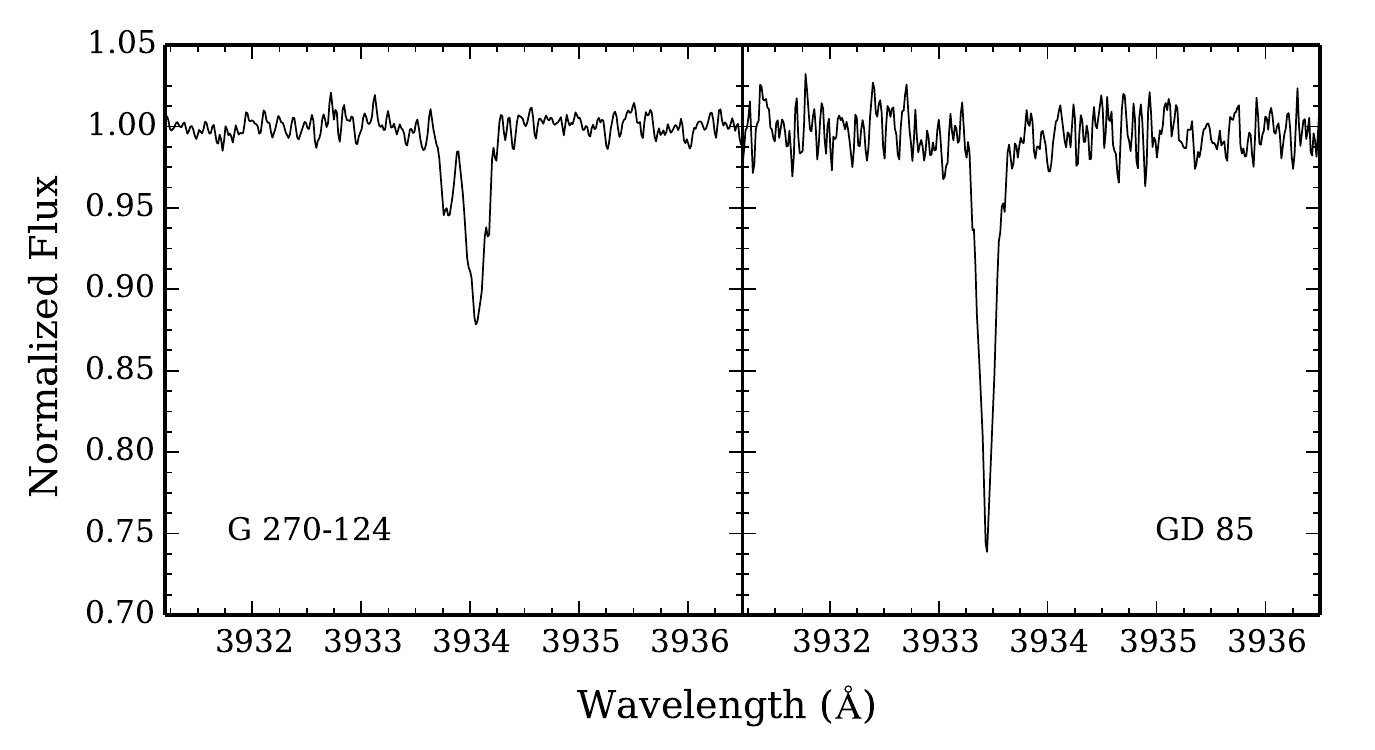}
\caption{HIRES spectra in the region of the Ca K-line for G270-124 and GD 85. Wavelengths are in air and the heliocentric frame of rest.}
\label{fig:CaK-plot2}
\end{figure}

For stars that already have Ca K-line EW and/or abundance measurements derived from high resolution spectra in the literature, we adopt those values as indicated by references (2) to (8) in Table \ref{tab:params}.  
Otherwise, we present new Ca K-line equivalent width measurements (or upper limits) and abundance fits (or upper limit estimates) for nearly half of the sample stars, denoted by references (1*) and (1). While the presence of Ca K-lines in GD 408, GD 61, GD 303, G200$-$39, and GD 378  has been long known \citep{kenyon1988, sion1988}, our updated high-resolution EW measurements provide a significant increase in precision over those discovery papers.  In addition, we perform new fits based on our HIRES data for the [Ca/He] abundances for GD 408, GD 303, and GD 378.  GD 378 has a rich HIRES spectrum, and a comprehensive abundance analysis will be presented separately (Klein et al., in prep.).  Ton 345's high-Z absorption lines, discovered along with its calcium infrared triplet gas emission lines \citep{gaensicke2008}, have been analyzed using high-resolution data \citep{jura2015,wilson2015}, but its Ca K-line EW measurement is first recorded here.   Finally, we report new Ca K-line detections and abundance measurements for G270$-$124 and GD 85 (Figure \ref{fig:CaK-plot2}).

The fit abundances are extracted using an atmosphere code similar to that described in \cite{dufour2007}. For a given star, the effective temperature, surface gravity and H abundance are fixed to the values given in \cite{rolland2018}, and the Ca abundance is varied in steps of 0.5 dex.  This grid of model atmospheres and synthetic spectra are used to interpolate a fit to the final abundance by employing a $\chi^2$ minimization algorithm to find the abundance that yields the best fit to the line.

\subsection{G270-124}
The Ca K-line feature in G270-124 has two components, an EW = 8 m{\AA} line with radial velocity of 10 km s$^{-1}$ and a somewhat stronger Ca K-line of 27 m{\AA} at a radial velocity of 31 km s$^{-1}$ (left panel of Figure \ref{fig:CaK-plot2}). The larger, more red-shifted component is close to the (photospheric) He5876 line velocity of 36 km s$^{-1}$, suggesting its origin is likely photospheric, while the weaker component is clearly not photospheric and may be either interstellar or circumstellar (further consideration of these possibilities is beyond the scope of this text and deferred to future work).  G270$-$124 is already known to have other heavy elements in its atmosphere from an analysis of {\it FUSE} data carried out by \citet{desharnais2008}, where they reported pollution by carbon, silicon and iron, as well as non-photospheric absorption lines of N, C and O.  Quoting a relative velocity accuracy for {\it FUSE} of $\pm$6 km s$^{-1}$, \citet{desharnais2008} measured the relative velocity between photospheric and ISM lines to be 36 km s$^{-1}$. This is larger than the 21.2 km s$^{-1}$ velocity difference between the two 
Ca K-lines of our HIRES spectrum, so these relative velocity measurements may or may not be reconcilable, and we conclude that it is uncertain whether the blue-shifted Ca K-line has a similar origin as the N, C, and O blue-shifted ground state lines in the {\it FUSE} data.   Nonetheless, our derived abundance for calcium from the stronger (atmospheric) Ca K-line is [Ca/He] = -8.08. Comparing that to the Si and Fe abundances from \citet{desharnais2008} of -7.5 and -7.2, respectively, corresponds to abundance ratios of Ca/Si = 0.26 and Ca/Fe = 0.13 by number, which are well within the range of values observed for a variety of WDs polluted by rocky material \citep{jurayoung2014}.  All these factors taken together, it is highly likely that the stronger K-line detection comes from the WD photosphere along with the other heavy elements.

\subsection{GD 85}
 \citet{sion1988} previously noted a hint of a Ca K-line feature in GD 85 and placed an upper limit of EW $<$ 100 m{\AA}.  Our HIRES data clearly detect the Ca K-line in GD 85 with EW = 50$\pm$4 m{\AA}, for which we derive an abundance of [Ca/He] = -8.80.  The radial velocities of the Ca K-line and He5876 agree to within 2 km $^{-1}$.  Given the radial velocity agreement, together with the moderately strong equivalent width, there is little doubt that GD 85's Ca K-line originates from the WD's atmosphere.

\subsection{Stars with UV Carbon Detections \label{sec:carbon}}
At least a quarter of our sample stars have published photospheric carbon absorption in their ultraviolet spectra (identified in Table \ref{tab:core-data}). Six of those also have Ca pollution in addition to other high-Z lines (GD 408, G270-124, GD 40, Ton 345, G200-39, and GD 378, references from Table \ref{tab:params} and \citet{desharnais2008}).  The carbon in these stars likely originates from planetary accretion along with the other heavy elements. In these cases the He5876 core profiles appear to be governed by the overall pollution and H abundances, as discussed in the remainder of the paper.  

Five stars have C detected in the UV, but no other high-Z elements, they are: LAWD 74 \citep[= LDS 678B = EGGR 131,][]{wegner1981}, L791-40 \citep[= LTT 9491,][]{koester1982iue, wegner1983}, BPM 17088 \citep[= HE 0308-5635,][]{petitclerc2005}, GD 190 \citep{provencal2000}, and GD 358 \citep{sion1989}.  The first two are at the cool end of our sample with $T_{\rm eff}$ $\sim$10,500\,K, and the last three are at the hot end with $T_{\rm eff}$ = 22,000 $-$ 25,000\,K.  The presence of atmospheric carbon in the cool stars is understood as the result of convective dredge-up \citep{koester1982, pelletier1986, dufour2005}, while the origin of carbon in the hot DBs is as yet unclear \citep[e.g.][]{brassard2007, koester2014db, dunlapclemens2015, koesterkepler2019}. In this special subset of sample stars with only C pollution, the C abundance ranges from $-$6.5 $<$ [C/He] $<$ -5.5. Thus, even though these stars do not display Ca or any other heavy elements, their atmospheres do contain a potentially significant abundance of high-Z material.  However, we also note that the effective temperatures of this set are all far outside the $T_{\rm eff}$ range (14,000\,K $-$ 17,500\,K) where we see the inversion cores.  We consider both of these conditions in Section \ref{sec:model}.
\\

We note that some of the stars in Table \ref{tab:params} have been observed with the Cosmic Origins Spectrograph (COS) on the {\it Hubble Space Telescope} ({\it HST}) under various program IDs.  A full analysis of complementary HIRES and COS spectra of stars from {\it HST} program ID 13453 (PI: M. Jura) will be treated in a separate paper (Klein et al., in prep). It is possible that future analyses will yield somewhat different elemental abundances than those used here in Table \ref{tab:params}, but we note that all anticipated changes will not qualitatively impact the conclusions of the current paper.

\begin{deluxetable}{lrrcc}
\tabletypesize{\small}
\tablecaption{He5876 Line Core Data \label{tab:core-data} }
\tablehead{
     \colhead{Name} & \colhead{[H/He]} &\colhead{[Ca/He]} & \colhead{He5876} & \colhead{Inv. EW} \\
         \colhead{}	  & 	\colhead{}	& \colhead{detected} &   \colhead{core}  & \colhead{(m{\AA})} 
}
\startdata
G26-10	&	$<$ $-$6.49	&	$<$ $-$11.1	&	inv	&	175	$\pm$	17	\\
GD 406	&	$-$6.47	&	$<$ $-$10.8	&	inv	&	194	$\pm$	30	\\
GD 124	&	$<$ $-$6.46	&	$<$ $-$10.9	&	inv	&	218	$\pm$	49	\\
G241-6	&	$-$6.43	&	$-$7.25	&	abs	&				\\
GD 325	&	$-$6.37	&	$<$ $-$10.6	&	inv	&	112	$\pm$	23	\\
L 182-61	&	$<$ $-$6.32	&	$<$ $-$10.3	&	inv	&	88	$\pm$	9	\\
PG 1144-085	&	$<$ $-$6.32	&	$<$ $-$10.1	&	inv	&	115	$\pm$	31	\\
G256-18	&	$<$ $-$6.28	&	$<$ $-$10.7	&	inv	&	128	$\pm$	13	\\
PG 1411+219	&	$-$6.26	&	$<$ $-$10.8	&	inv	&	123	$\pm$	27	\\
G26-31	&	$<$ $-$6.22	&	$-$8.6	&	inv	&	49	$\pm$	6	\\
GD 40\tablenotemark{$\dag$}	&	$-$6.14	&	$-$6.88	&	abs	&				\\
G64-43	&	$-$6.08	&	$<$ $-$10.6	&	inv	&	151	$\pm$	15	\\
GD 85	&	$<$ $-$5.99	&	$-$8.80	&	inv	&	25	$\pm$	9	\\
EGGR 133	&	$-$5.97	&	$<$ $-$10.1	&	inv	&	35	$\pm$	15	\\
GD 408\tablenotemark{$\dag$}	&	$-$5.95	&	$-$9.59	&	inv	&	86	$\pm$	9	\\
PG 1644+199	&	$-$5.68	&	$-$10.3	&	inv	&	77	$\pm$	10	\\
KUV 05034+1445	&	$-$5.46	&	$<$ $-$10.7	&	inv	&	31	$\pm$	9	\\
\hline											
HE 2334-4127	&	$-$5.34	&	$<$ $-$8.9	&	abs	&				\\
GD 303	&	$<$ $-$5.34	&	$-$7.8	&	abs	&				\\
G119-47	&	$-$5.33	&	$<$ $-$11.9	&	abs	&				\\
L791-40\tablenotemark{$\dag$}	&	$-$5.27	&	$<$ $-$12.0	&	abs	&				\\
LAWD 74\tablenotemark{$\dag$}	&	$-$5.16	&	$<$ $-$12.5	&	abs	&				\\
G270-124\tablenotemark{$\dag$}	&	$-$5.14	&	$-$8.08	&	abs	&				\\
Ton 345\tablenotemark{$\dag$}	&	$-$5.10	&	$-$5.95	&	abs	&				\\
GD 190\tablenotemark{$\dag$}	&	$<$ $-$4.84	&	$<$ $-$7.6	&	abs	&				\\
EC 12522-2855	&	$<$ $-$4.82	&	$<$ $-$7.8	&	abs	&				\\
BPM 17088\tablenotemark{$\dag$}	&	$<$ $-$4.82	&	$<$ $-$7.8	&	abs	&				\\
PG 1326-037	&	$<$ $-$4.81	&	$<$ $-$8.0	&	abs	&				\\
LAWD 90	&	$-$4.78	&	$<$ $-$9.6	&	abs	&				\\
LP475-242	&	$-$4.68	&	$-$9.2	&	inv	&	53	$\pm$	16	\\
Feige 4	&	$-$4.63	&	$<$ $-$9.1	&	abs	&				\\
PG 2354+159	&	$<$ $-$4.59	&	$-$8.1	&	abs	&				\\
GD 358\tablenotemark{$\dag$}	&	$<$ $-$4.58	&	$<$ $-$7.7	&	abs	&				\\
GD 378\tablenotemark{$\dag$}	&	$-$4.45	&	$-$8.35	&	abs	&				\\
GD 243	&	$-$4.35	&	$<$ $-$9.1	&	abs	&				\\
G200-39\tablenotemark{$\dag$}	&	$-$4.26	&	$-$9.3	&	abs	&				\\
WD 1916-362	&	$-$4.22	&	$<$ $-$7.6	&	abs	&				\\
GD 61	&	$-$4.21	&	$-$7.9	&	abs	&				\\
HE 0110-5630	&	$-$4.18	&	$-$7.9	&	abs	&				\\
GD 205	&	$-$4.07	&	$-$7.9	&	abs	&				\\
\enddata
\tablenotetext{$\dag$}{Atmospheric carbon detected in the UV, see Sections \ref{sec:carbon} and \ref{sec:model}}
\tablecomments{List of stars from Table \ref{tab:params} ordered by [H/He].  The appearance of the He5876 core is either an inversion (inv, Figure \ref{fig:lowZ}), or absorption (abs, Figure \ref{fig:highZ}).  A crossover between the populations occurs around [H/He] = -5.4.}

\end{deluxetable}

\begin{figure}
\includegraphics[width=85mm]{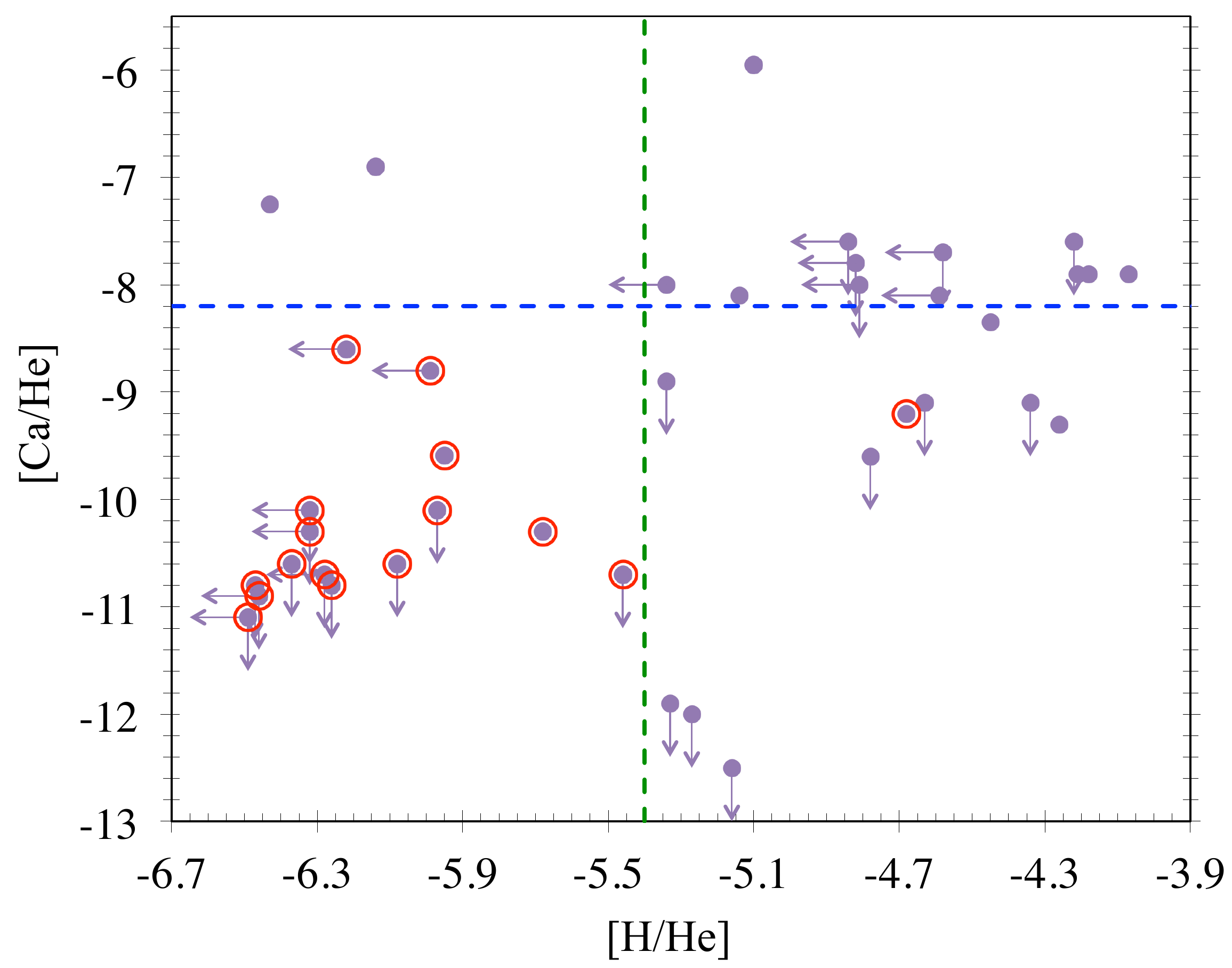}
\caption{Abundances and upper limits (arrows) from Table \ref{tab:core-data}. Red circles denote stars with a core inversion. The green and blue dashed lines indicate crossover values of hydrogen and calcium abundances (respectively), in agreement with where the core profiles of Figures \ref{fig:models_h} and \ref{fig:models_metals} transition from inversion to absorption. All of the stars with a core inversion have both low [H/He] and low [Ca/He], only with the exception of LP475-242 which has a relatively higher H abundance; at this point we do not have an explanation for this outlier (see also, Section \ref{sec:profiles}). }
\label{fig:Ca-He_space}
\end{figure}
 
\begin{figure}
\includegraphics[width=85mm]{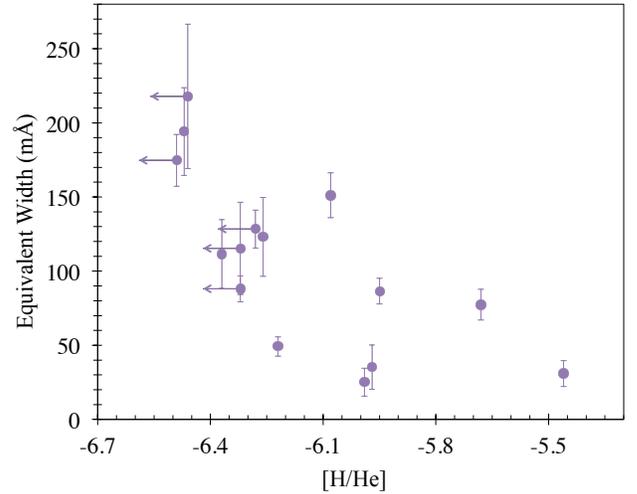}
\caption{Stars from the lower left quadrant of Figure \ref{fig:Ca-He_space} and their Equivalent Widths from Table \ref{tab:core-data}.  There is a correspondence between the strength of the He5876 core emission and the H abundance in the atmosphere. A similar plot (not shown) with [Ca/He] on the abscissa follows the same trend.}
\label{fig:emission_strength}
\end{figure}

\section{He 5876 line core profiles\label{sec:profiles}}
He5876 core inversion is present for more than ${1 \over 3}$ of our DB sample, while a single absorption core is observed for stars with a relatively high abundance of heavy elements and/or hydrogen. Core inversions are not seen in other strong He lines (e.g. He~{\small I} $\lambda$4472), which is consistent with our models as shown in Section \ref{sec:model}.  Intermediate temperature single WDs generally do not exhibit line core emission in their spectra, but an exception is PG 1225$-$079 which displays core emission in its Ca H- \& K-lines \citep{klein2011}. However, those features are most likely a unique detection of chromospheric activity in this WD \citep[][Section 3.1]{dufour2012}.

Both when a core inversion is seen and when a single absorption core is seen, the central absorption core is often narrow (Figures \ref{fig:lowZ} and \ref{fig:highZ}). Narrow He5876 absorption cores have been previously noticed in three cool ($\sim$11,000 K) helium-dominated and heavily-polluted WDs: GD 362, GD 16, and PG 1225$-$079 \citep{zuckerman2007, klein2011}. In hydrogen-dominated atmosphere (DA) white dwarfs, narrow non-LTE cores in H$\alpha$ and H$\beta$ have also been observed at high-resolution \citep{reid1996, falcon2010, zuckerman2013}. We note, however, that our proposed explanation for the presence of narrow absorption cores in our DB sample does not require non-LTE effects. Rather, it is likely due to the lower density/temperature (and thus reduced broadening) in the upper region of the atmosphere (Rosseland mean optical depth ($\tau_R$) $\lesssim 10^{-6}$, see Figures \ref{fig:models_h} and \ref{fig:models_metals}) where the line core is formed.

Having assembled the available parameters for the dataset $-$ $T_{\rm eff}$, log $g$, [H/He], [Ca/He] $-$ we can examine trends and correlations.  Is it the presence of hydrogen, or of high-Z elements, that is responsible for these distinctive core features?  We believe it is a combination of the two.  We see that in Table \ref{tab:core-data}, which is ordered by increasing [H/He] abundance, there is a transition in the core profile appearance at around [H/He] = $-$5.4.  Stars with [H/He] less than that (above the line in Table \ref{tab:core-data}) nearly all display an inversion core.  The exceptions are the two heavily polluted WDs, G241-6 and GD 40, which have [Ca/He] $>$ $-$7.5. Referring to Figure \ref{fig:Ca-He_space}, observationally, it appears the criterion for having an inverted He5876 core is an H abundance less than $-$5.4, as well as a Ca abundance less than $\sim$ $-$8.  We believe this is most likely due to the additional opacity introduced by the hydrogen and pollutants, and these limits are well matched by our model analysis in Section \ref{sec:model}.

Examination of the set of spectra with inverted He5876 cores reveals a range in the degree of prominence of the emission feature. We parameterize the strength of the inversion by measuring the EW of the emission feature as follows:  We chose the ``continuum'' to be set by the feature endpoints as the lowest points on the blue and red wings of the line core profile (ignoring the self-reversed central dip which in a few cases is somewhat lower than the minimum of the line wings).  Relative to that baseline, we use IRAF's {\it splot ``e''} mode to measure the EW, which is calculated from the summed flux of the pixels between the feature endpoints above the baseline. For each star, we performed the measurement three times for a varied choice of continuum endpoints according to variations in the noise, and then estimated an uncertainty from the standard deviation of those separate measurements. We also tried extracting EWs by Voigt fitting the two emission peaks (using IRAF {\it splot ``d''} mode), which returned reasonable fits.  But it is more challenging to get a total EW from that since the fit functions overlap.  Values for the EWs from the {\it splot ``e''} flux summing method are listed in Table \ref{tab:core-data}.  As shown in Figure \ref{fig:emission_strength}, the strength of the core emission decreases as the total number abundance of hydrogen increases, in excellent agreement with the model predictions shown in Figure \ref{fig:models_h}.   

A note about the outlier from Table \ref{tab:core-data} and Figure \ref{fig:Ca-He_space}, LP~475$-$242 (which happens to be a Hyades member), that displays a self-reversed emission core despite its relatively high hydrogen abundance ($\log\,\rm H/He=-4.68$).  We considered that one possible explanation could be that the source is a double degenerate, such that most or all the hydrogen line absorption does not come from the dominant DB atmosphere. Following \citet{bedard2017}, we performed a variety of spectroscopic fits using either DB(A)+DA or DBA+DB(A) composite models, but we found that either the fits to the H or He lines are poor, and/or the overall flux is inconsistent with photometry.  The result is that a double degenerate solution appears unlikely, as the source is always better fit with a single DBA model.  This agrees with the lack of radial velocity variations over three epochs of Keck/HIRES observations, which are all broadly consistent with a radial velocity of $\sim85 \pm 2$ km s$^{-1}$ \citep[][plus one unpublished epoch $-$ PI Jura]{reid1996, zuckerman2013}.

\section{MODEL ATMOSPHERE ANALYSIS\label{sec:model}}

\begin{figure*}
  \centering
  \includegraphics[width=2\columnwidth]{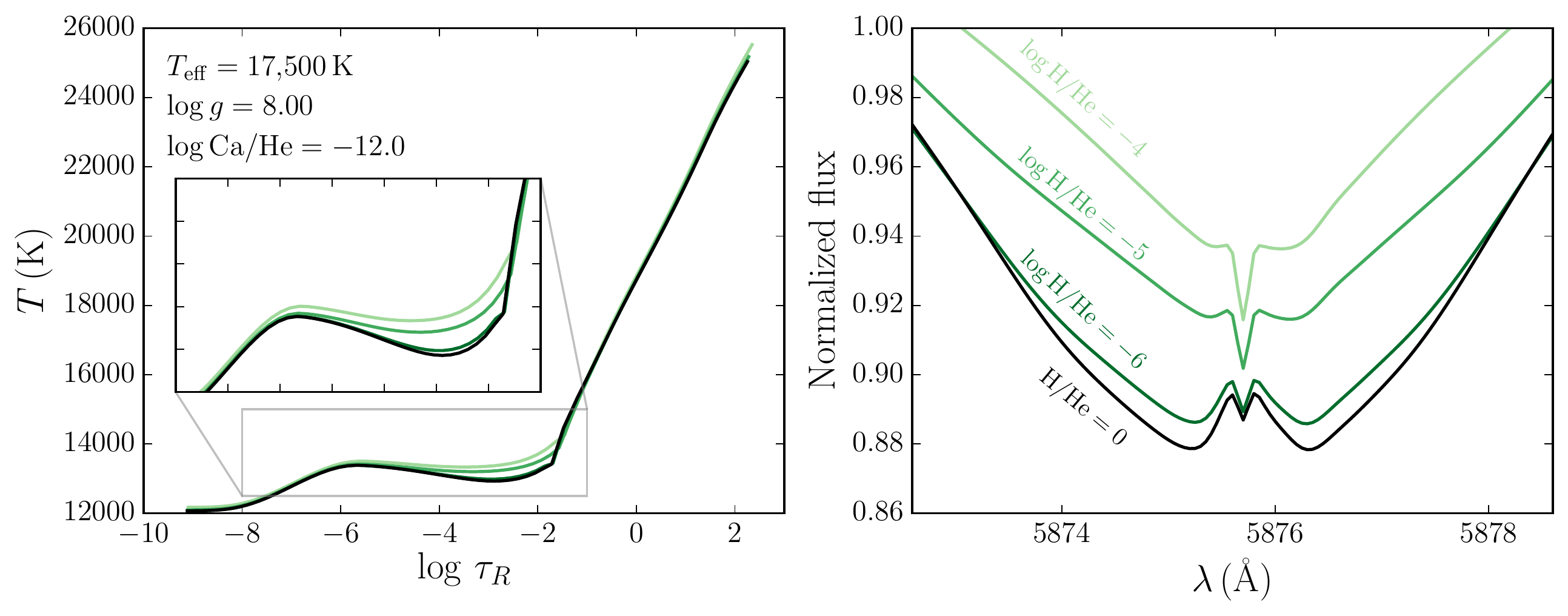}
  \caption{$T(\tau_R)$ structure (i.e., temperature as a function of the Rosseland mean optical depth) of DB(A)Z models with different H abundances (left) and corresponding 
    He~{\small I} $\lambda$5876 core profiles (right).}
  \label{fig:models_h}
\end{figure*}

\begin{figure*}
  \centering
  \includegraphics[width=2\columnwidth]{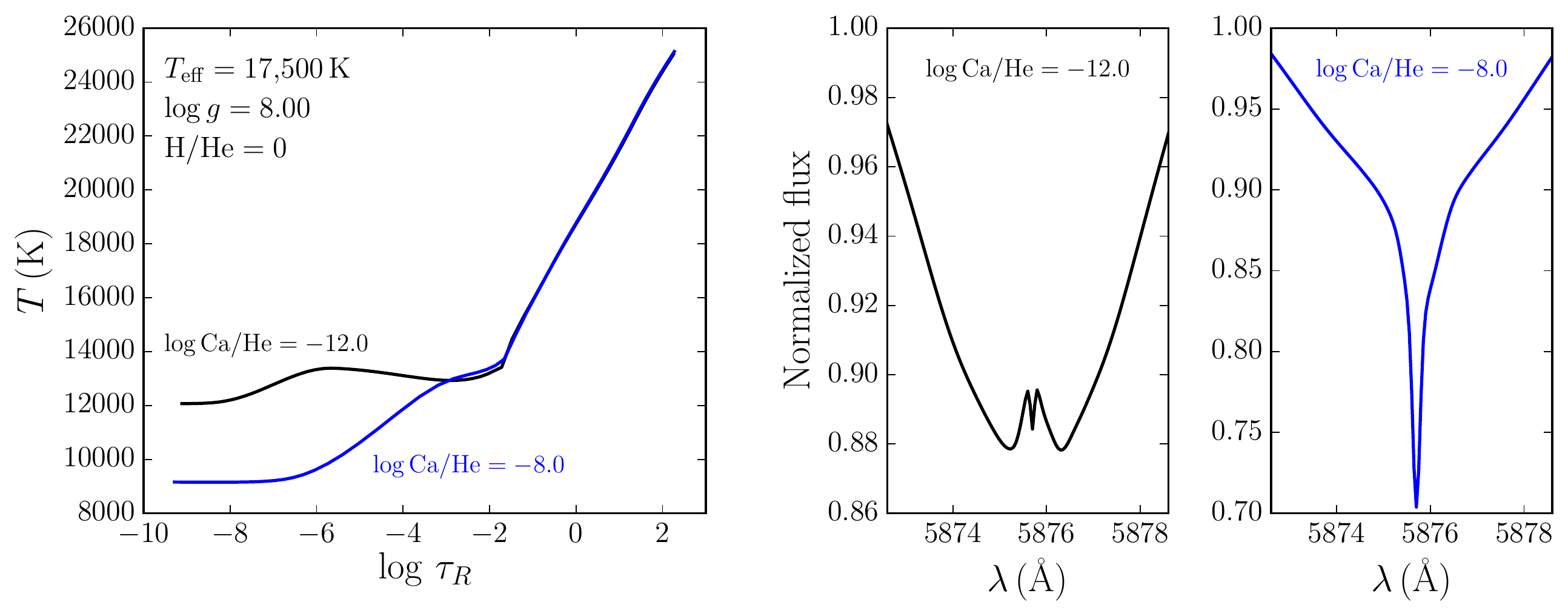}
  \caption{$T(\tau_R)$ structure of DBZ models with different high-Z abundances (left) and corresponding He~{\small I} $\lambda$5876 core profiles (right). The abundance ratios of heavy elements are scaled to the abundance of Ca to match the abundance ratios of CI chondrites \citep{lodders2003}.}
  \label{fig:models_metals}
\end{figure*}

We now attempt to explain the distinctive He5876 core profiles described in the previous
section with detailed model atmospheres. Using the atmosphere code described in \cite{dufour2007} and \cite{bergeron2011},
we computed a grid of DB(AZ) models to investigate whether the core inversion profile can be seen in synthetic spectra.
Our grid extends from $T_{\rm eff}=14{,}000$\,K to $20{,}000$\,K and includes ${\rm H/He}$ abundance ratios of 0, 
$10^{-6}$, $10^{-5}$ and $10^{-4}$, and ${\rm Ca/He}$ abundance ratios of 0, $10^{-12}$, $10^{-10}$ and $10^{-8}$. The model atmospheres extend from $\tau_R = 200$ all the way to $\tau_R = 10^{-9}$, which is important to properly capture the emission/absorption in the core region of He5876.
We found that for $17{,}000\,{\rm K} \leq T_{\rm eff} \leq 19{,}500$\,K and pure 
helium atmospheres, synthetic spectra do display a self-reversed emission 
structure similar to that described above, resulting from a double inversion in the temperature profile (i.e.~$T(\tau_{R})$) that coincides 
with the He5876 line-forming region.  
The temperature inversion and the resulting emission structure remain visible if the atmosphere contains a small quantity of H. However, the emission structure slowly disappears as the H/He abundance ratio is increased (Figure \ref{fig:models_h}). This is in excellent agreement with the observed relationship between the strength of the inversion feature and the level of H (Figure \ref{fig:emission_strength}).
Similarly, the emission structure is still visible if a small quantity of high-Z material is added to the atmosphere model, but disappears if the atmosphere is heavily polluted (Figure \ref{fig:models_metals}). We found that these transitions are due to the feedback of the additional hydrogen and/or high-Z lines' opacity on the temperature profile.  Overall, the models predict many features that are consistent with the observations:
\begin{itemize}
\item The shape of the emission feature is similar to that observed, with a $\approx$ 2\,{\AA} width in wavelength scale.
This indicates that the temperature inversions occur at the correct optical depths, as the width of the
emission structure is determined by the broadening of He5876 in those regions.
\item The emission structure is still visible if the atmosphere contains a small quantity of H, consistent
  with the results listed in Table~\ref{tab:core-data} and Figure \ref{fig:Ca-He_space}\footnote{With the exception of LP~475$-$242.},
 %We note that LP~475$-$242 is a possibleoutlier as it displays a self-reversed emission core despite its relatively high hydrogen abundance ($\log\,\rm H/He=-4.68$).}
 and the strength of the feature decreases with increasing H abundance, as observed in Figure \ref{fig:emission_strength}.
\item The emission feature is visible for objects that have no detected high-Z lines and for those that are
  lightly polluted, in agreement with Table~\ref{tab:core-data} and Figure \ref{fig:Ca-He_space}.
\item The emission structure gives way to a single absorption core if the atmosphere is heavily polluted with high-Z material
  (Figure~\ref{fig:models_metals}), also consistent with the trend identified in Table~\ref{tab:core-data} and Figure~\ref{fig:Ca-He_space}. This transition is due to the disappearance of the emission core following the inhibition of the temperature inversions.
\end{itemize}

\begin{figure}
  \centering
  \includegraphics[width=\columnwidth]{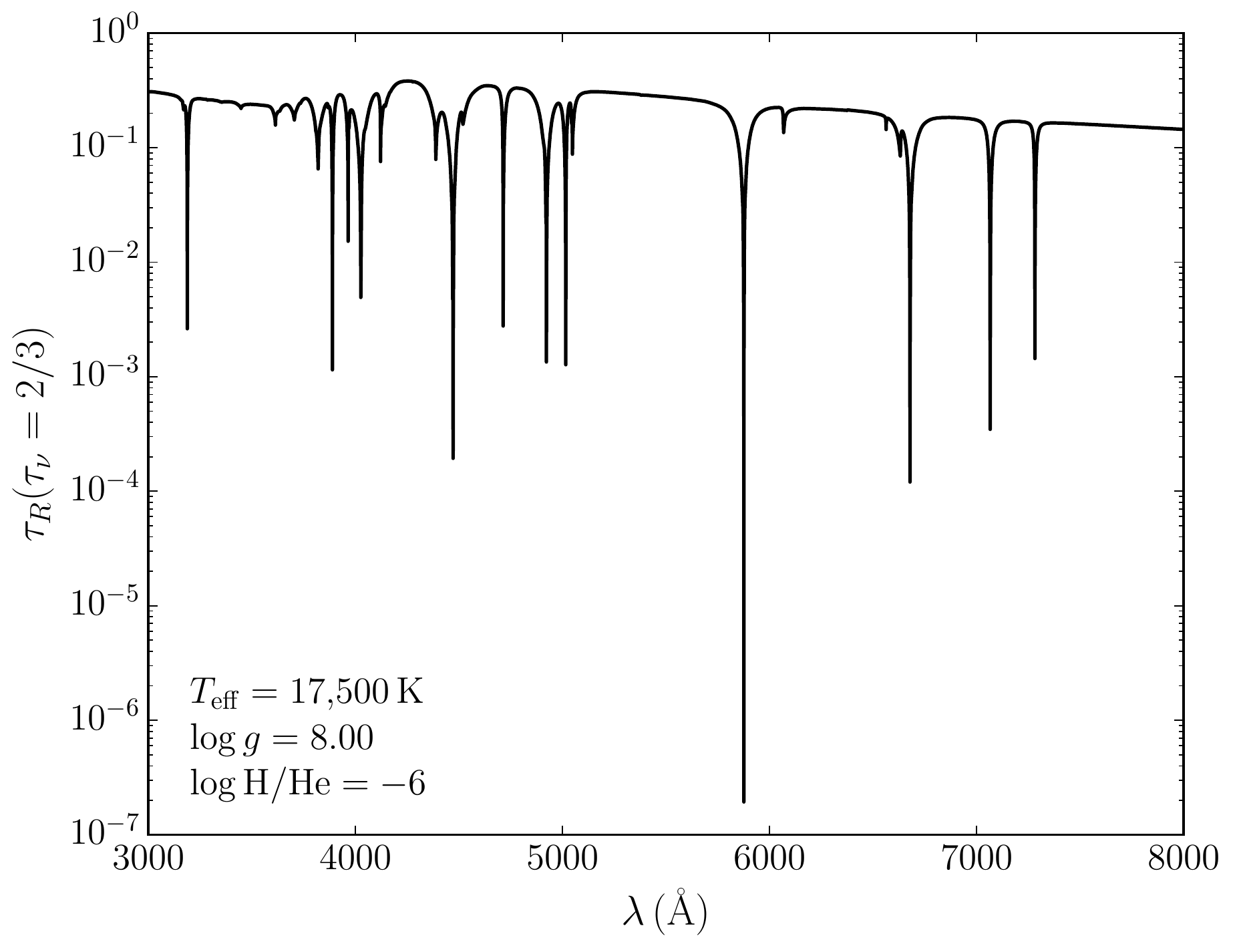}
  \caption{Rosseland mean optical depth at $\tau_{\nu}=2/3$ as a function of $\lambda$. Different atmospheric layers are probed at different wavelengths as the opacity changes. Only the core of He5876 is formed high enough in the atmosphere to be affected by the double inversion in the temperature structure (Figures~\ref{fig:models_h} and \ref{fig:models_metals}).}
  \label{fig:taunu}
\end{figure}

Moreover, our models can explain why only He5876 shows a self-reversed emission core (and not other He lines such as He {\small I} $\lambda$4472). Figure~\ref{fig:taunu} shows the Rosseland mean optical depth at $\tau_{\nu}=2/3$ as a function of $\lambda$. This type of figure is useful to visualize which atmospheric layers are probed at different wavelengths. In this case, it shows that only He5876 has a core formed high enough in the atmosphere to probe the double inversion in the temperature structure around $\tau_R \approx 10^{-6}$. The cores of all other He lines are formed deeper in the atmosphere and are thus not affected by the double inversion, which explains the nondetection of self-reversed emission cores for those lines.

Since all five objects for which only C was detected in the UV have an absorption core for He5876 (see Section \ref{sec:carbon} and Table \ref{tab:core-data}), we also checked how C pollution alone affects the emission feature. We found that the temperature inversion and the emission feature do disappear for models polluted by C, but the level of pollution required for the complete disappearance of the emission feature exceeds that measured in the five C-polluted stars of our sample. However, the effective temperatures of those objects are well below or above the temperature range where the self-reversed emission feature is observed (Table~\ref{tab:params}). This suggests that it is their temperature and not the C pollution that explains their He5876 core shape.

However, two major issues remain: (1) we do not see the core emission profile for models cooler 
than 17{,}000\,K, while
observations show that it exists down to $\approx$14{,}500\,K and (2) the models underestimate the amplitude of the emission features.  The temperature structures of cooler models
do contain a temperature inversion similar to that shown in Figures~\ref{fig:models_h} and \ref{fig:models_metals}, 
but its location does not coincide with the He5876 line-forming region.
A key piece of missing physics in our models
that could be at the origin of these problems is the 3D treatment of convection \citep[e.g.,][]{tremblay2013}. 
Our code uses the simplistic mixing length theory \citep[ML2/$\alpha$=1.25,][]{bergeron2011}.
Compared to this prescription of convection, an accurate 3D treatment can lead to important changes in the temperature 
structure \citep[for DB models, see Figure 4 of][]{cukanovaite2018}.  This could affect the location and/or strength of the temperature inversions and potentially resolve the temperature and emission strength
discrepancies identified above.  In that sense, the self-reversed emission core could prove to be a powerful diagnostic tool to calibrate the temperature structure of DB atmosphere models.  

Another missing piece of physics is non-LTE effects, which can be particularly important in the upper atmosphere. The inclusion of non-LTE effects in hot WDs ($>$25,000\,K) has been shown to have significant impact on the line profiles and temperature scales \citep[e.g.][]{napiwotzki1997, hubeny1999}. Thus it is possible that departures from LTE could be responsible for a shifted temperature scale and/or core emission strength, however it is unclear how big this effect can be in the cooler range 10,000\,K $<$ $T_{\rm eff}$ $<$ 25,000\,K of our sample.

\subsection{Stratified models}
\begin{figure*}
  \centering
  \includegraphics[width=2\columnwidth]{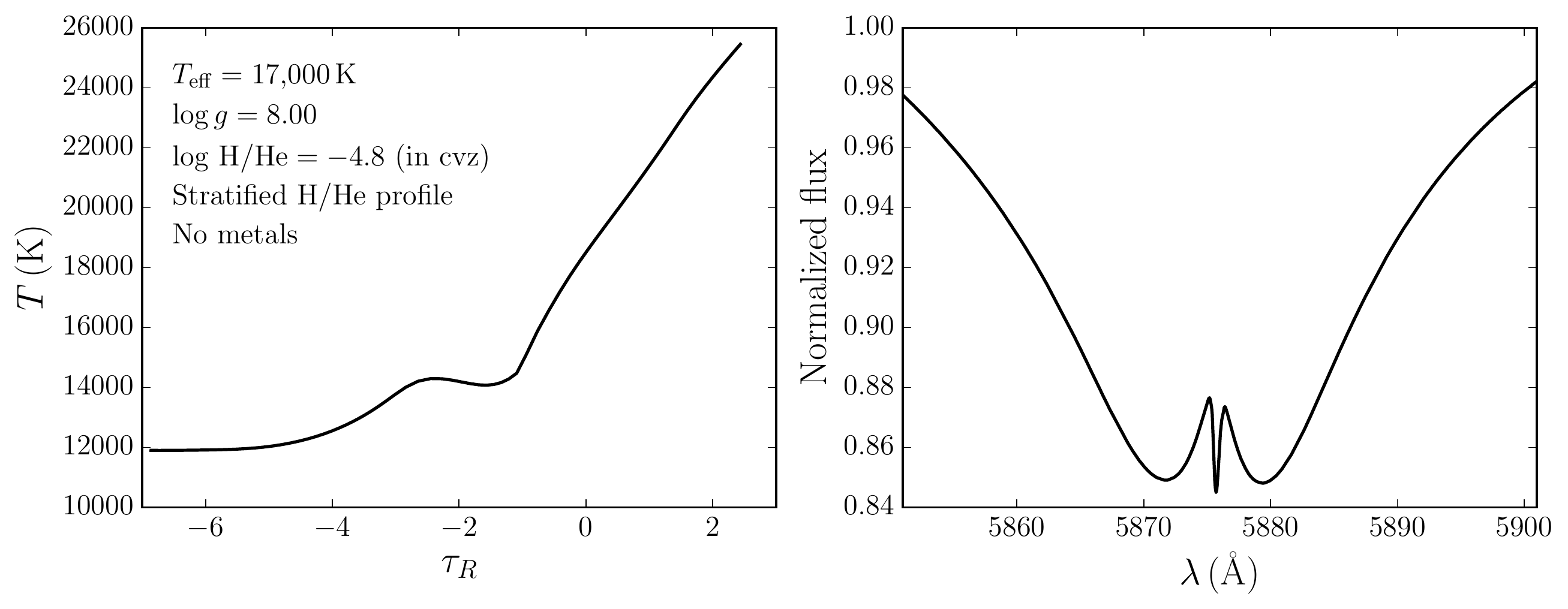}
  \caption{$T(\tau_R)$ structure and He~{\small I} $\lambda$5876 core profile of a mixed H/He atmosphere model where chemical 
    stratification is included assuming diffusive equilibrium. Note that the $\log\,{\rm H/He}$ ratio given in the left
    panel corresponds to the abundance in the convection zone (cvz) of the model.}
  \label{fig:inversion_and_triplet}
\end{figure*}

As most stars that show a self-reversed emission core
have mixed H/He atmospheres, we now investigate whether chemical stratification $-$ previously not included in our 
atmosphere models mentioned just above $-$ could help solve the temperature discrepancy.
Mixed H/He white dwarf atmospheres are expected to be chemically stratified, with a thin H
layer floating in diffusive equilibrium above the He envelope \citep{jordan1986,vennes1992,manseau2016}. 
Diffusive equilibrium can safely be assumed as the diffusion timescale is much smaller than the white dwarf cooling time.
To investigate whether this stratification could play a role in the emission structure observed for He {\small I} 
$\lambda$5876, we modified our atmosphere code to include 
the stratification of H/He. To do so, we assume a constant H/He ratio in the convective layers of the atmosphere
models and we impose diffusive equilibrium for the radiative layers. The diffusion equilibrium profile is
obtained from Equation~5 of \cite{pelletier1986},
which gives the relative diffusion velocity of two species with different atomic masses $A_i$
and average charges $\bar{Z}_i$ \citep[see also][Equation 1]{vennes1988}. This diffusion equation includes the contributions of
the concentration gradient, gravity and the electric field.\footnote{Note that both thermal diffusion
\citep{paquette1986} and radiative forces \citep{fontaine1979,chayer1995} can be safely neglected                          
for the objects considered in this work.} Assuming diffusive equilibrium, we obtain,
\begin{equation}
  \begin{split}
    \frac{\partial \ln c_{\rm He}}{\partial r} = 
    &\left( \frac{A_{\rm H} \bar{Z}_{\rm He} - A_{\rm He} \bar{Z}_{\rm H}}{\bar{Z}_{\rm H} + y \bar{Z}_{\rm He}}\right) \frac{m_p g}{kT} + \\
    &\left( \frac{\bar{Z}_{\rm He} - \bar{Z}_{\rm H}}{\bar{Z}_{\rm H} + y \bar{Z}_{\rm He}} \right) \frac{\partial \ln p_i}{\partial r},   
    \label{eq:diffusion}
  \end{split}
\end{equation}
where $c_{\rm He} = n_{\rm He} / (n_{\rm He} + n_{\rm H})$, $y=n_{\rm He}/n_{\rm H}$, $m_p$ is the proton mass, 
$g$ is the surface gravity, $k$ is the Boltzmann constant and $p_i$ is the ionic pressure ($p_i = p_{\rm H} + p_{\rm He}$).
Equation \ref{eq:diffusion} is solved for $y(r)$ at every iteration of the temperature correction
procedure \citep[for details on how to solve Equation \ref{eq:diffusion}, see][Appendix~A]{genestphd}, so that
a self-consistent structure is obtained once the model has converged. 

We computed a grid of models with effective temperatures ranging from $14{,}000$\,K to $20{,}000$\,K and H/He abundances in
the convection zone (cvz) from $10^{-7}$ to $10^{-4}$.
We find that for a narrow domain of atmospheric parameters ([H/He] $\geq -5$ and
$T_{\rm eff} \geq 16{,}500\,{\rm K}$) a double inversion appears in the temperature structure and leads
to the formation of a self-reversed emission core for He~{\small I} $\lambda$5876 (an example of such a model is shown in
Figure \ref{fig:inversion_and_triplet}). We conclude that including chemical stratification in our models does 
not lead to a better agreement with the observations:
\begin{itemize}
\item We are still unable to obtain an emission profile for stars between 14{,}500 and 16{,}500\,K.
\item The hydrogen abundance needed to see an emission profile is not compatible with the observations.
  We find that we need [H/He] $\geq -5$ in the convection zone, which means even more H at the photosphere
  and in the line-forming region of the atmosphere. This leads to H$\alpha$ profiles that are much stronger than those observed in our sample and opposes the [H/He] trend of Table \ref{tab:core-data}. 
\item The shape of the emission profile obtained from our stratified models is much wider ($\approx 8\,{\rm \AA}$) than what is actually observed ($\approx 2\,{\rm \AA}$).
\item Over the range of $T_{\rm eff}$ and [H/He]) where a self-reversed emission core is predicted for He5876, the stratified models also predict emission in the He {\small I} $\lambda$4472 line core, but that is not observed in any of our sample data.
\end{itemize}

Overall, it seems that chemical stratification is not the solution we are looking for. However, we refrain from completely
ruling out stratification as an explanation for these observations, since our H/He stratification profiles could be off due to
the fact that we do not include 3D effects. In particular, we neglect convective overshoot,
which means that we  underestimate the height that the convective eddies can reach. This necessarily implies that 
our chemical stratification profile is different from reality (overshoot would extend the region of the atmosphere where
chemical homogeneity is assumed), although it is not clear by how much.

\section{CONCLUSIONS\label{sec:conclusions}}

The analysis of a brightness-selected sample of DB WDs observed at high-resolution demonstrates a correspondence between the hydrogen abundance and the degree of atmospheric pollution by heavy elements  with the formation of the He {\small I} 5876 line core. Our model atmospheres go a long way in explaining this phenomenon as due to the presence, or suppression, of temperature inversions in the upper atmosphere, however the effective temperature range for the phenomenon and strength of the emission are not well matched.  This may be due to potentially important physical mechanisms that are not included in our current models, i.e.~the effects of 3D atmosphere calculations, convective overshoot, and/or non-LTE effects, which may change the temperature structure and/or the H/He stratification. In view of these considerations, a direct comparison of models with observations remains an aim of future work.
In any event, the He5876 core profile structure appears to be a new and powerful diagnostic tool for the calibration of temperature profiles in DB atmospheres, providing a useful check for how closely a given WD atmosphere model resembles reality. Since WD models are fundamental for deriving abundances of accreted exoplanets in polluted WDs, this calibration tool may also have implications for exoplanetary studies.

\section{ACKNOWLEDGEMENTS}
We thank the anonymous referee for useful suggestions that improved the manuscript. We are also thankful for helpful conversations with Boris G{\"a}nsicke and Detlev Koester that have led to clarifications and improvements.
This work has been supported by grants from NASA and the NSF to UCLA.  B.K.~acknowledges support from the APS M. Hildred Blewett Fellowship,  S.B.~acknowledges support from the Laboratory Directed Research and Development program 
of Los Alamos National Laboratory under project number 20190624PRD2.  D.R. acknowledges support from the UCLA-SRI summer program for undergraduate research.  S.X. is supported by the international Gemini Observatory, a program of NSF's NOIRLab, which is managed by the Association of Universities for Research in Astronomy (AURA) under a cooperative agreement with the National Science Foundation, on behalf of the Gemini partnership of Argentina, Brazil, Canada, Chile, the Republic of Korea, and the United States of America. The majority of the data presented herein were obtained at the W.M. Keck Observatory, which is operated as a scientific partnership among the California Institute of Technology, the University of California and the National Aeronautics and Space Administration. The Observatory was made possible by the generous financial support of the W.M. Keck Foundation.  We recognize and acknowledge the very significant cultural role and reverence that the summit of Mauna Kea has always had within the indigenous Hawaiian community.  We are most fortunate to have the opportunity to conduct observations from this mountain.   Additionally, some of the data are based on observations collected at the European Southern Observatory under ESO programme 165.H-0588.  This research has made use of the services of the ESO Science Archive Facility, SIMBAD, SAO/NASA ADS, the Montreal White Dwarf Database, IRAF, Python, Matplotlib.

\end{CJK*}
\bibliographystyle{apj}
\bibliography{apj-jour,BKrefs}

\end{document}